# Optically controlling the competition between spin-flips and intersite spin transfer in a Heusler half-metal on sub-100 fs timescales

Sinéad A. Ryan[1][†]*, Peter C. Johnsen[1][†], Mohamed F. Elhanoty[2][†], Anya Grafov[1], Na Li[1], Anna Delin[3,4,5], Anastasios Markou[6,7], Edouard Lesne[7], Claudia Felser[7], Olle Eriksson[2,5], Henry C. Kapteyn[1,8], Oscar Grånäs[2], and Margaret M. Murnane[1]

[†]Equal contributors

*Corresponding author: sinead.ryan@colorado.edu

[1]JILA, University of Colorado Boulder, 440 UCB, Boulder, CO 80309, USA

[2]Division of Materials Theory, Department of Physics and Astronomy, Uppsala University, Box-516, SE 75120, Sweden

[3]Department of Applied Physics, School of Engineering Sciences, KTH Royal Institute of Technology, AlbaNova University Center, SE-10691 Stockholm, Sweden

[4]Swedish e-Science Research Center (SeRC), KTH Royal Institute of Technology, SE-10044 Stockholm, Sweden

[5]Wallenberg Initiative Materials Science for Sustainability, Uppsala University, 75121 Uppsala, Sweden

[6]Physics Department, University of Ioannina, 45110 Ioannina, Greece

[7]Max Planck Institute for Chemical Physics of Solids, 01187 Dresden, Germany

[8]KMLabs Inc., Boulder, CO, USA.

## Abstract

The direct manipulation of spins via light may provide a path toward ultrafast energy-efficient devices. However, distinguishing the microscopic processes that can occur during ultrafast laser excitation in magnetic alloys is challenging. Here, we study the Heusler compound $Co_2MnGa$, a material that exhibits very strong light-induced spin transfers across the entire M-edge. By combining the element-specificity of extreme ultraviolet high harmonic probes with time-dependent density functional theory, we disentangle the competition between three ultrafast light-induced processes that occur in $Co_2MnGa$: same-site Co-Co spin transfer, intersite Co-Mn spin transfer, and ultrafast spin-flips mediated by spin-orbit coupling. By measuring the dynamic magnetic asymmetry across the entire M-edges of the two magnetic sublattices involved, we uncover the relative dominance of these processes at different probe energy regions and times during the laser pulse. Our combined approach enables a comprehensive microscopic interpretation of laser-induced magnetization dynamics on timescales shorter than 100 fs.

## Introduction

Developing the next generation of spintronic devices will require a new level of manipulation of complex materials and their spin states at short timescales. However, fully exploiting these capabilities for more energy-efficient nanotechnologies requires a detailed understanding of the physical mechanisms underlying nanoscale spin manipulation(*1*). The interaction of ultrafast laser pulses with magnetic materials can induce complex spin dynamics, both during and after the laser pulse(*2–4*). When combined with ultrafast extreme UV and soft X-ray probes, it is possible to detect element-specific spin dynamics in multi-component magnetic systems(*5*, *6*), providing rich new information not accessible using visible light. Initial studies of laser-induced spin manipulation assumed that changes to the magnetic state were a secondary process triggered by an initial hot electron distribution. In this hot electron model, electrons first absorb laser photons during a femtosecond laser excitation pulse. This is followed by electron-phonon mediated spin-flips and other scattering processes to absorb the angular momentum and demagnetize the sample on timescales of approximately 0.5 ps(*7*) or longer.

More recent studies that probe the instantaneous magnetization of different elements have shown that much faster manipulations of spins are possible using light, on femtosecond and even attosecond timescales. In one finding(*8*), a new transient magnetic state was observed in laser-excited Ni, where a magnetic phase transition is launched within a laser pulse (<20 fs), provided the electron temperature exceeds the Curie temperature, $T_c$. In a second finding(*9*, *10*), light-induced spin transfer within the laser pulse duration was observed between two elements in the same material (i.e. intersite). This intersite spin transfer behavior was predicted theoretically(*11*, *12*) and observed experimentally in Heusler compounds(*9*, *12*) and ferromagnetic alloys(*10*, *13*). Nevertheless, several unanswered questions about the underlying microscopic processes remain: what constitutes a clear signature of intersite spin transfer, and how do we distinguish it from other ultrafast effects such as spin-flips, electron redistribution and demagnetization?

Here we address these challenges by combining the element-specificity of extreme ultraviolet (EUV) high harmonic probes with time-dependent density functional theory (TD-DFT) and detect a definitive signature of light-induced intersite transfer of spin polarization in a Heusler compound $Co_2MnGa$. This material can be grown in a highly-crystalline phase with a half-metallic bandgap (Fig. 1C) and thus might be expected to support a strong optical intersite spin transfer effect(*14*). We observe very strong enhancements of the magneto-optical signal across the entire M-edge due to light-induced spin transfer - a behavior never observed in any material to date. To identify specific excitation pathways, one needs to distinguish how same-site spin transfer, intersite transfer and ultrafast spin-flips mediated by spin-orbit coupling respectively change the magnetic moments of Mn and Co, and how these processes manifest themselves in the transient EUV magneto-optic signal. To achieve this goal, we scan the energy of an EUV probe in order to measure the spin dynamics across the entire M-edges of the two magnetic sublattices involved. Then, by comparing experimental observations with theory based on TD-DFT, we uncover the relative dominance of same-site Co-Co spin transfer (Fig. 1A), intersite Co-Mn transfer (Fig. 1A), and ultrafast spin-flips mediated by spin-orbit coupling (Fig. 1B). Moreover, the contributions from these different processes to the light-induced spin manipulation change as a function of time and laser intensity. Although theoretical studies have predicted the fluence dependence of intersite spin transfer(*11*, *15*), no experimental studies have been published to date. By changing the pump fluence and probe energy, we show that one can both identify and tune the competing microscopic mechanisms underlying light-induced spin manipulation on ultrafast timescales (<100 fs). The combination of detailed theoretical insight, excellent sample quality, and an extensive experimental dataset has allowed us to demonstrate side-by-side theoretical and experimental comparisons of ultrafast spin dynamics in complex magnetic alloys.

**Spin Transfer in Magnetic Alloys**

It was recently predicted that optical spin pumping from one metallic sublattice to a second sublattice can transiently enhance the magnetic moment of the second metallic sublattice within the laser pump pulse(*11*, *14*), in a process often called optical intersite spin transfer (OISTR). Since the origin of the OISTR effect is optical excitation from occupied to unoccupied states, this presents the potential for few-femtosecond optical manipulation of spin states. Moreover, it might be expected that OISTR has a particularly strong signatures in Heusler compounds because of their unique band structure(*14*).  Heuslers are a group of ordered magnetic compounds with a chemical formula of $X_2YZ$ for full Heuslers, or XYZ for half-Heuslers(*16*). This class of compounds is particularly exciting as it supports a wide range of materials with excellent chemical stability, and with electronic and magnetic properties that can be engineered based on the number of valence electrons of the constituent elements. Heusler compounds host a number of remarkable ground states that include topological insulators(*17–20*), half-metals(*21*) and superconductors(*22*, *23*), and are promising candidates for technological applications such as thermoelectric(*24–27*) and spintronic devices(*28–32*). The magneto-optical properties of Heuslers have been of great interest for the past 40 years following the measurement of the largest visible Magneto-Optical Kerr Effect (MOKE) signal on MnPtSb in 1983(*33*). In addition, half-metallicity (where one spin channel is gapped at the Fermi level while the other is partially filled and hence conducting) was first detected in a Heusler material(*34*), and recent investigations have identified non-quasiparticle states in a Heusler metal(*35*). More specifically for the present investigation, Heuslers have been the focus of many ultrafast magnetism studies. Furthermore, the half-metallicity can lead to unique responses to optical pumping since the available excitation channels are spin selective(*14*). The ultrafast MOKE responses of Heuslers and other half-metals were first demonstrated using visible probes(*12*, *36–38*).

Experimentally, OISTR was previously investigated using visible MOKE in Heuslers(*12*), followed by measurements with element-specific extreme ultraviolet (EUV) high harmonic probes in  the Heusler compound $Co_2MnGe$(*9*) as well as FeNi alloys(*10*). These initial experiments were followed by L-edge measurements indicating transient ferromagnetism in an antiferromagnet(*39*). However, in all of these studies, the OISTR-like enhancement was only observed at *one specific probing energy,* and the corresponding absorption edge was only probed at one(*9*, *39*) or two(*10*) energies. In other studies, OISTR was inferred from the respective demagnetization rates of Co and Pt in a CoPt alloy(*13*) or Ni in a stack of Ni/Pt multilayers(*40*).  However, in both of these cases no enhancements of the magnetic circular dichroism (MCD) were predicted or observed.

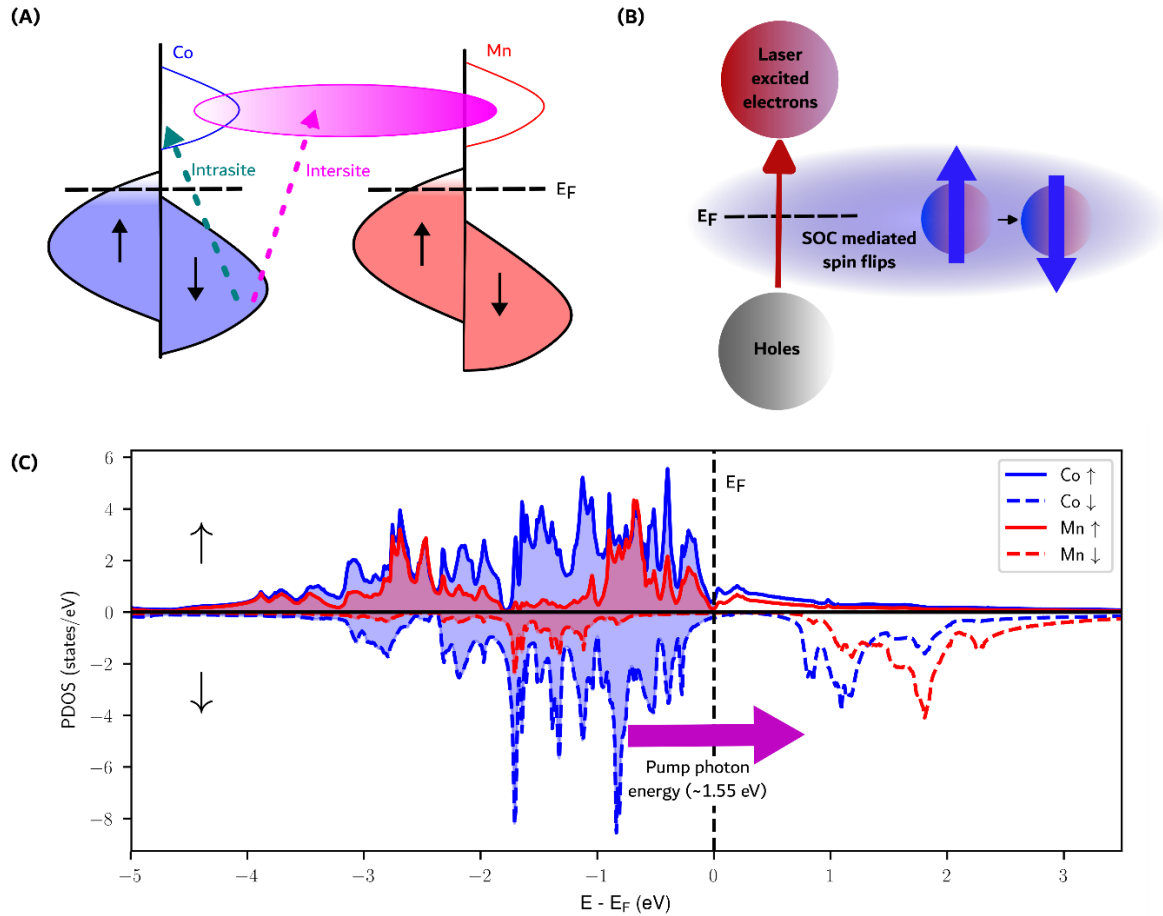

**FIG. 1. The three excitation types included in the theoretical calculations, and the ground state density of states. (A)** Optical excitations within the same species (intrasite, green dotted line) which are predominantly Co-Co minority band excitations, as well as excitations between different species (intersite, pink dotted line). The pink shaded area indicates a final state composed of a hybridized band with predominantly Mn character. **(B)** A spin-flip mediated by spin-orbit coupling (SOC) that is induced by the laser-excited non-equilibrium charge distribution. Such spin-flips contribute to demagnetization. **(C)** Calculated d-shell density of states and their population (PDOS) with spin and element specificity. PDOS in an occupation weighted projected density of states as outlined in the Materials and Methods section. The black dotted line is the Fermi Energy ($E_F$). The y-axis is divided into two, with the positive axis containing majority spin states and the negative axis containing minority spin states (signified with black arrows, representing spin direction). The energy of the pump laser's photons is represented with a pink arrow. Ga atoms contribute very few states at the Fermi energy (not shown).

OISTR is not the only effect that can cause a transient enhancement in the magneto-optical signal. When a femtosecond laser pulse excites a material, the electron population is redistributed during and after the pulse. This can lead to shifts and broadening of absorption edges on the same timescales as the laser pump pulse. These effects have been observed in experimental studies of Ni at the M-edge using circular dichroism(*41*, *42*) and transient absorption spectroscopy(*43–45*), as well as theoretical works studying Ni(*46*) with circular dichroism at the L-edge. These effects can lead to signatures that appear as transient enhancements in the MOKE or circular dichroism signal at specific probe energies – despite the fact that there is no overall increase in magnetization of the sample(*41*). Another study(*47*) showed vastly different Co demagnetization rates above and below the Co L-edge in a [Co/Pd] multilayer structure which were attributed to energy dependent spin-flip rates. Thus, it is critical to implement unambiguous measurements and combine with theory in order to capture the true signature of OISTR and distinguish it from local changes in the magneto-optical signal due to charge redistribution or spin-flips. We show below that this can be achieved by measuring the magnetic asymmetry across the full absorption edges on the two magnetic sublattices involved in the intersite spin transfer, and then comparing the measurements to the static and transient asymmetry from TD-DFT.

**Results**

Measurements were made using the X-MATTER beamline(*48*) in Boulder on a 20nm-thick epitaxial $Co_2MnGa$ film (see Materials and Methods for details). A 40-55 fs, ~800 nm laser pump pulse is used to excite the sample. The resulting dynamics were probed using extreme ultraviolet (EUV) light generated through high harmonic generation (HHG), with a pulse duration of ~25 fs. The transverse magneto-optical Kerr effect (TMOKE) is used to probe the magnetic asymmetry. See Fig. S1 in the Supplementary Materials for a schematic.

In Fig. 2, we present the full energy resolved $Co_2MnGa$ asymmetry in both experiment and theory. An incident fluence of 3.4 mJ/cm$^2$ is used corresponding to an absorbed fluence of 2.2 mJ/cm$^2$. This absorbed fluence value (2.2 mJ/cm$^2$) was used in the TD-DFT calculations. We measure and model the changes following laser excitation to determine how excitations manifest themselves across the Mn and Co M-edges. These results are shown in Figs. 2A and 2C, where we plot the experimental and theoretical asymmetries in the ground state as well as at 80 fs following laser excitation.

The magnetic asymmetry at the Mn M-edge peak resonance (~47-51 eV) shows a transient reduction in the TMOKE signal, as shown in Fig. 2A. Experimental data which include more time points for Mn energies are available in the Supplementary Materials (Fig. S2). Given the density of states of $Co_2MnGa$ (Fig. 1C), and our excitation photon energy of ~1.55 eV, we attribute the transient reduction of the Mn TMOKE signal to ultrafast demagnetization processes as well as the transfer of minority band electrons from Co, both of which reduce the spin polarization of Mn. The transient reduction at the Mn-edge at 80 fs is also reproduced in the calculations shown in Fig. 2C, where it can be seen that the experiment and theory agree very well, both for the static ground state as well as for the driven system. The calculations based on TD-DFT demonstrate ultrafast demagnetization by spin-flips driven by the spin-orbit coupling (SOC)(*49*). SOC mixes the spin states so that spin is no longer a pure quantum number(*50*). Unlike dipole excitations, spin-flip transitions do not conserve the total magnetization of the sample and will have a net demagnetizing effect.

The TMOKE asymmetry signal is strongest when the probe is resonant with the energy difference between the 3p core states and the Fermi energy. For this reason, the signals at the Co and Mn peaks predominantly arise from excitations to valence states which are in the close vicinity of the

Fermi energy. Across the Co-edge (~57-72 eV), we observe a transient enhancement. However, the size of the enhancement depends strongly on the probing EUV photon energy, as shown in Fig. 2B. The pump-induced changes at two different energies (at the M-edge, and above the edge) display maximal enhancements of approximately 5% and 14%, respectively. The changes in the TD-DFT calculated asymmetry at similar energy regions reveal a similar trend, as shown in Fig. 2D. Indeed, the experimental and theoretical data that can be compared in Figs. 2B and 2D demonstrate very similar features, both when it comes to the time- and energy dependence of the transient optical asymmetry, as well as the general shape of the measured and calculated data. The different strengths of enhancements at different probing energies can be understood by considering that there is strong optical pumping that moves spin-polarized electrons in Co from lower to higher energy bands. The transient Co signal at any given energy is sensitive to this redistribution of the electron population, despite the fact that the overall spin polarization of Co is unchanged by these intraspecies (same-site) optical excitations. Furthermore, the strength of the demagnetizing spin-flips is also energy dependent. The strongest spin-flip excitation occurs near the peak of the signal (60.4 eV) as will be discussed in more detail below.

We note that although Fig. 2D depicts theoretical transient enhancements close to 30% at 61.5 eV, the calculated change in the Co moment is only 2.5%, as shown in Fig. 3C. It is clear that the small change in the Co magnetic moment is not the only process influencing the energy-dependent transient EUV TMOKE signals. It is noteworthy that the present investigation is the first to measure an OISTR effect across two entire absorption edges (Co and Mn), which allows for a much deeper analysis compared to studies with a limited number of energy probes (*9*, *10*, *39*). This allows us to conclude that the measured transient enhancement of the magnetic asymmetry signal is extremely energy dependent and can be large at specific probing energies. For example, near the zero-crossing of the Co asymmetry (59.3 eV), the magnitude of the signal is very sensitive to changes in the asymmetry shape, and an enhancement of greater than 100% is observed. Similarly, probing at 71.5 eV where the energy is far above the Co peak and the absolute signal is small, one may observe transient enhancements in the signal of over 2000%, see Fig. S3 in the Supplementary Materials.

The pump pulse controls the OISTR effect(*15*) by modifying the electron population around the Fermi energy. This modification spans an energy range of approximately twice the pump photon energy (or ~3.1 eV). To understand the potential OISTR effects within the sample, we analyze the available states above and below the Fermi energy. Fig. 1C displays the ground state density of states (DOS) of $Co_2MnGa$, where one can identify the insulating gap in the spin down channel that gives rise to the half-metallic character. Below the Fermi energy, 3d-shell minority spins in Co provide more occupied electronic states available for pump excitation than states in Mn. In contrast, Mn has more available unoccupied states just above the Fermi energy. The probabilities of individual transitions come from the dipole transitions accessible by the pump laser, as analyzed in detail in Ref. (*9*). The TD-DFT calculations access these excitations by solving the time-dependent Kohn-Sham Hamiltonian (Eq. (1) in Materials and Methods). Based on the excitations and depletions shown in Figs. 3A and 3B, that are analyzed in detail below, we infer that excitations from Co minority states to Mn minority states are more probable than other types of interspecies excitations.

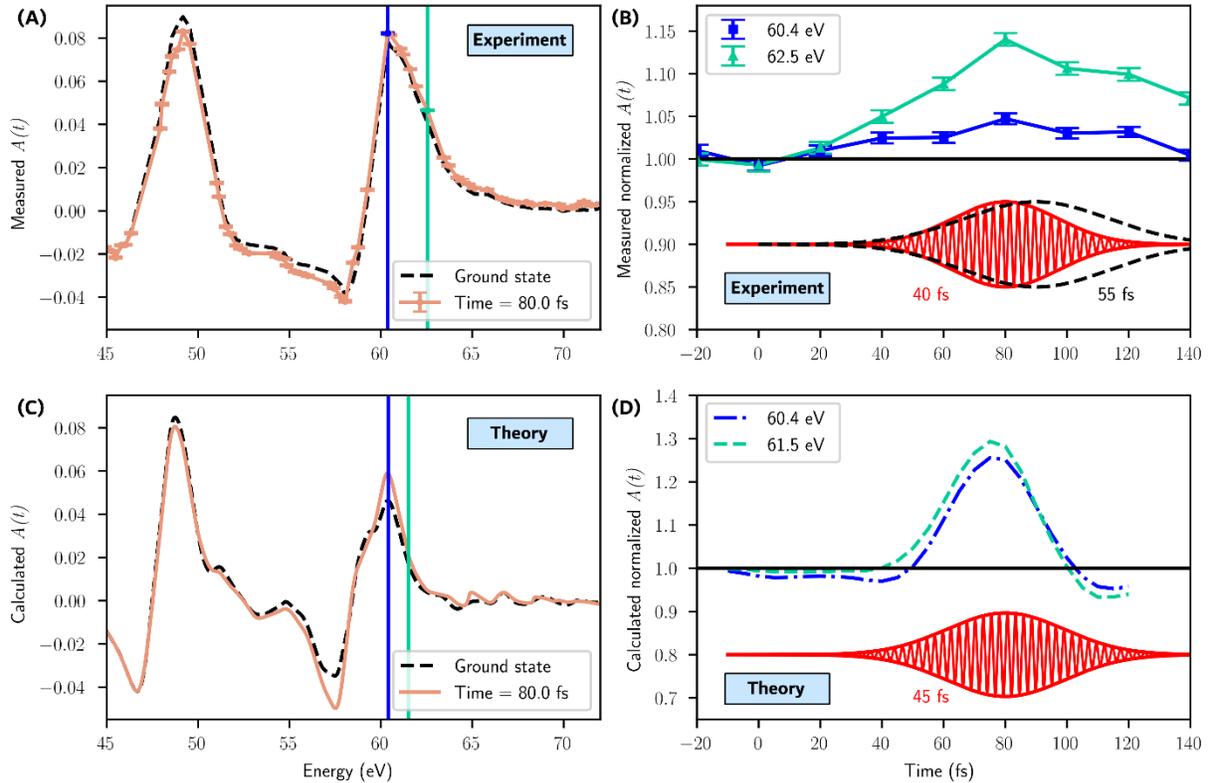

**FIG. 2. Experimental and theoretical magnetic asymmetries. (A)** The experimentally measured asymmetry in the ground state and at 80 fs following a 2.2 mJ/cm$^2$ laser excitation. **(B)** The measured transient asymmetry signals at selected energies (marked with colored vertical lines in A) normalized by their ground state values. A depiction of the experimental pump pulse appears below the data with an arbitrary vertical scaling. The experimental pulse is represented as an oscillating electric field with a full width half maximum (FWHM) that was measured to range from 40-55 fs. **(C)** The theoretically calculated asymmetry in the ground state and at 80 fs. **(D)** The simulated transient asymmetry signals at selected energies (marked with colored vertical lines in C) normalized by their ground state values. To simulate the measurement probe, the theoretical dynamics were convolved with the intensity profile corresponding to a 25 fs FWHM Gaussian electric field envelope. The unconvolved data appear in the Supplementary Materials (Fig. S8). The theoretically modelled pump electric field, 45 fs FWHM, appears below the data with an arbitrary vertical scaling.

Moving minority spins from Co to Mn leads to an enhancement of the Co magnetization accompanied by a simultaneous reduction in the Mn magnetization. This is consistent with the asymmetry changes measured and modeled in Figs. 2A and 2C as well as the changes in magnetic moment simulated in Fig. 3C.

The origin of the transient changes in the theoretical and experimental magnetic asymmetry curves can be understood through the changes in the occupation of the 3d states as depicted in Fig. 3. In Figs. 3A and 3B, we show a snapshot of the changes in the energy resolved majority and minority spin occupations for the 3d states at 70 fs for Co and Mn, respectively. This figure should be interpreted as follows: a negative signal below the Fermi level indicates a depletion of electron states at the corresponding energy, while a positive signal above the Fermi level indicates the occupation of electron states that were empty in the ground state. As the figure shows, some of the occupied majority and minority spin states below the Fermi level become depleted. In addition, the previously empty states in the minority and majority spin channels above the ground state Fermi energy become partially filled.

Several processes contribute to the behavior seen in Fig. 3: Co-Co spin pumping during the pump pulse, Co-Mn spin pumping during the pump pulse, and spin-flip excitations which exist both during and after the pulse excitation. The largest occupation increases and depletions are seen in the Co minority channel. From this we infer that Co-Co minority spin pumping is strong. The quantity of Mn-Mn and Mn-Co transitions is low due to a lack of available initial and final states with an appropriate energy separation that would be accessible to the laser pump pulse, as shown in Fig. 1C. By integrating the total change in spin polarization across all energies (Figs. 3A and 3B) we obtain the net change in moment for each element (Fig. 3C). The magnetic moment of Co is maximally increased by 2.5% while the Mn moment simultaneously decreases by 2%. The theory predicts a transient reduction in the total number of Co minority spins accompanied by an increase in Mn minority spins. Therefore, we infer that the main OISTR pathway in this material is Co to Mn minority spin pumping. The calculated ground state moments for this sample are 2.72 $\mu_B$ per Mn atom, 0.77 $\mu_B$ per Co atom, 0.06 $\mu_B$ per Ga atom and an interstitial moment (not associated with any specific element) of -0.04 $\mu_B$ per unit cell. We note that while there are twice as many Co atoms as Mn atoms, the Mn atoms carry a magnetic moment more than three times as large as Co. This means that although the calculated percentage enhancement in Co is larger than the corresponding percentage decrease in Mn, the sample exhibits a net demagnetization.

To separate the fingerprints of the SOC-mediated spin-flips and spin transfer processes in the simulated dynamics, we make use of the time-varying amplitude of the simulated laser pulse. As the strength of the laser pulse changes, the signatures of each process appear in specific time and energy windows within the TMOKE spectra. To demonstrate this, we plot the theoretical asymmetry curves at 40 fs, 80 fs, and 100 fs, see Supplementary Materials Fig. S4.

At early times (e.g., the first 50 fs), while the incident electric field from the pump laser is still weak, the dynamics are dominated by SOC-mediated spin-flips. The laser excites electrons from regions located around the atomic nuclei (i.e. muffin-tin regions) to the interstitial region between atoms. In the interstitial regions, electrons are more delocalized, and spin-orbit effects are weaker (*49*). The excitation of electrons from the muffin tin regions to the interstitial regions creates a non-equilibrium distribution of electrons. This non-equilibrium distribution induces spin-flips in the regions with the strongest spin-orbit coupling, i.e. electrons near the Fermi energy in the muffin tin region (*49, 51*). Because this effect is strongest close to the Fermi energy, the spin-flips manifest themselves as a reduction in the intensity of the asymmetry signal, seen most strongly at the Co and Mn peaks (~60.4 eV and ~49 eV). The spin-flip signature in the asymmetry peaks is apparent in times between 0 – 50 fs, see Supplementary Materials Fig. S4. These spin-flips account for the reduction in Co magnetic moment seen in Fig. 3C on the same timescale.

At subsequent times (50-80 fs), as the incident electric field of the pulse grows, spin transfers start to outcompete the spin-flips. The spin transfers manifest themselves as an enhancement of the asymmetry across the entire Co-edge, and a small reduction of asymmetry at the Mn-edge, see Supplementary Materials Fig. S4. The small size of the reduction at the Mn peak is consistent with the small OISTR induced moment change (2.0%) as depicted in Fig. 3C. Optical excitations in the spin minority channel of Mn are mostly suppressed due to the optical gap, as shown in Fig.1C. After the maximum of the pump laser has passed, e.g. after 80 fs (see Supplementary Materials Fig. S7), the calculated TMOKE spectra show a relaxation of the electrons that were excited by spin transfers, and we therefore see a decay in the asymmetry at the Co-edge. After the pump pulse excitation ends, spin-flips dominate once again.

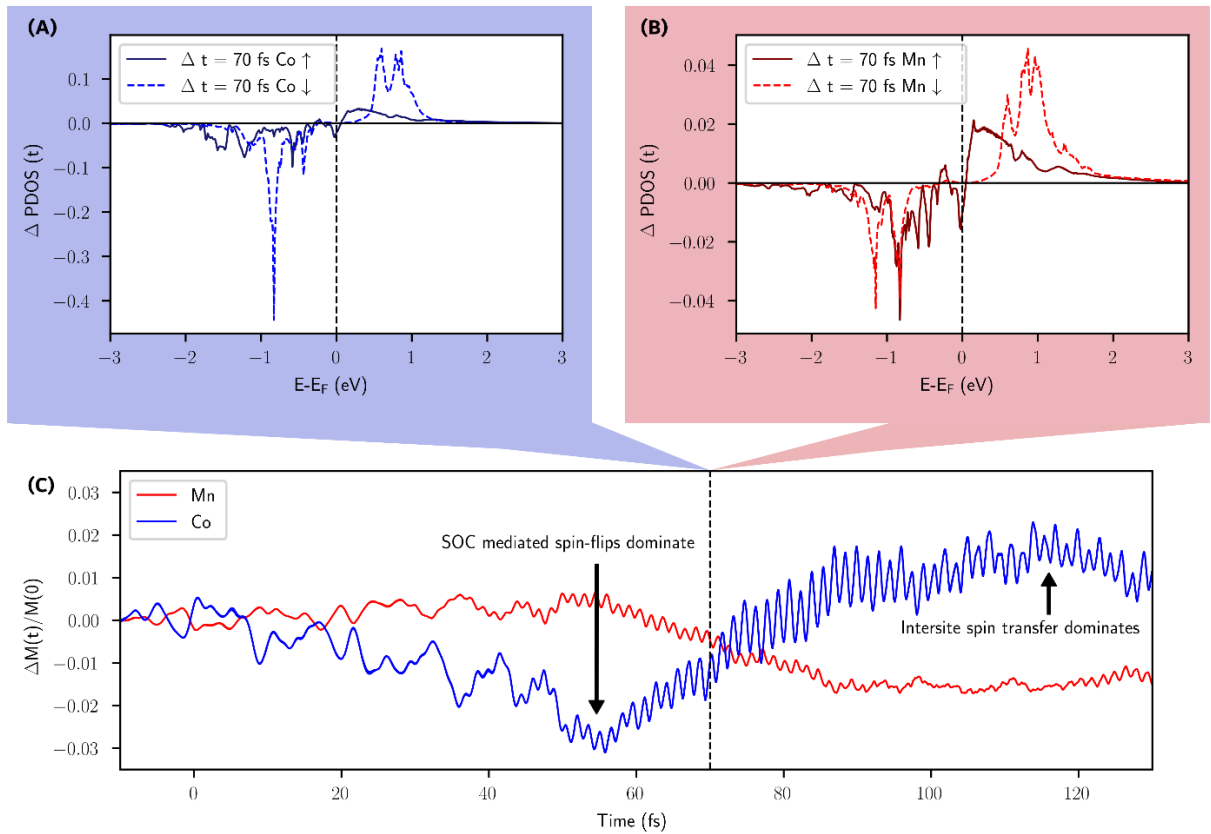

**FIG. 3. Theory calculations for pump induced excitations of Co and Mn. (A)** The energy-dependent and spin-specific change in the Co occupation 70 fs after laser excitation. The calculation of the occupation weighted projected density of states (PDOS) is explained in detail in the Materials and Methods section. A negative value signifies a depletion of electrons compared with the ground state and a positive value signifies an increase. **(B)** The same treatment is applied to Mn. Note the difference in scale for the Co and Mn PDOS. **(C)** The transient change in the total magnetic moments of Co and Mn following laser excitation. This is calculated by integrating the change in spin polarization across all energies shown in subfigures A and B. Initially, the response of Co is dominated by spin orbit mediated spin-flips. This is followed by a dominance of intersite spin transfer.

The dominance of spin-flips at early timescales that is predicted by theory (shown in Fig. 3C) appears only very weakly in the experimental data of Fig. 2B. This is attributed to a smearing of the response by the 25 fs EUV probe pulse. Indeed, when the theoretical data is convolved with the EUV probe pulse (as shown in Fig. 2D), we also do not observe strongly negative signals at early timescales. However, for the unconvolved data which appears in the Supplementary Materials, Fig. S8, negative values at early times are clearly observed.

At the Co asymmetry peak, there is strong competition between optical excitations, OISTR, and spin-flip effects. The MOKE signal enhancement at the resonance peak is diminished in comparison with the off-resonance regions where the OISTR and optical excitations dominate, as shown in Figs. 2B and 2D. This competition is experimentally exemplified in Fig. 4 where we show the fluence dependence at the Co peak (60.4 eV). These results show the first experimentally measured fluence dependence of OISTR.

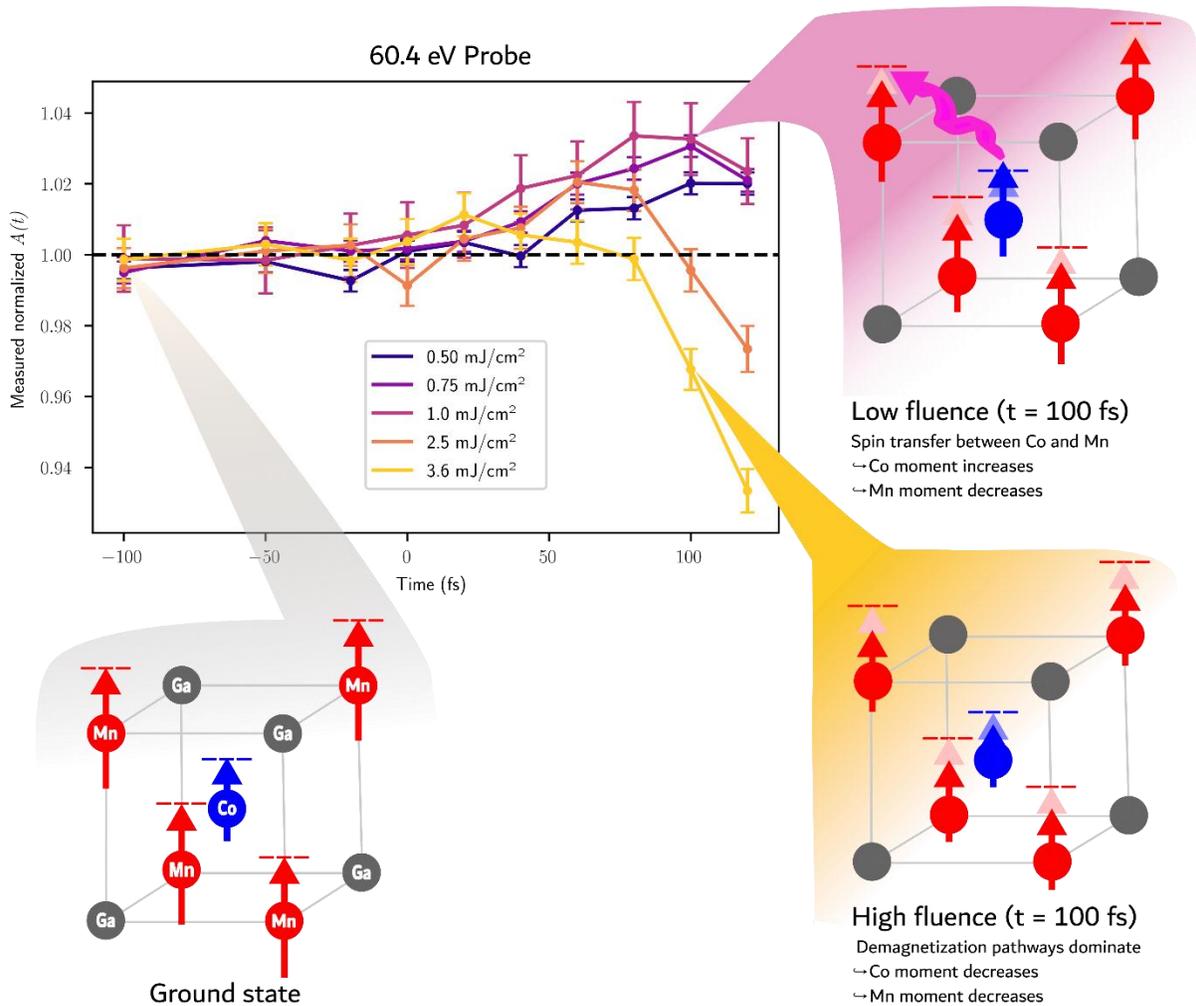

**FIG. 4. The fluence dependence of the MOKE asymmetry signal measured at the Co peak**. The Co peak is probed using 60.4 eV photons. The magnitude of the transient enhancement increases with increasing fluence up to 1.0 mJ/cm$^2$. Above 1.0 mJ/cm$^2$ it is diminished due to competition between demagnetization pathways and OISTR effects. The listed fluences are the absorbed fluences.

Theoretical predictions(15) indicate that the strength of the OISTR effect should be proportional to the fluence of the pump laser. At low fluences (<1.0 mJ/cm$^2$) at the Co peak, the transient enhancement of the TMOKE signal increases as the fluence increases (see Fig. 4). However, as the pump fluence is increased above 1.0 mJ/cm$^2$, the transient enhancement becomes smaller and the peak occurs at earlier times. We attribute this to the increasing dominance of demagnetization mechanisms at higher fluences – which begin to compete with, and overcome the OISTR effects. This is in contrast to the pre-edge and post-edge regions where optical excitations dominate over spin-flips (see Supplementary Materials Fig. S5). However, the TD-DFT calculations predict that the enhancement should continue increasing with fluence (see Supplementary Materials Fig. S6). For example, with an 8.4 mJ/cm$^2$ pump fluence, an OISTR enhancement of the Co moment of 10% is predicted, compared to only 2.5% for a fluence of 2.2 mJ/cm$^2$. Thus, the theoretical model used here underestimates the strength of the demagnetizing effects at the peak. This could be due to the fact that the theory only includes spin-flips and cannot model other demagnetizing effects such as magnon generation, electron-phonon coupling or superdiffusive spin currents, all of which would increase in strength with increasing pump-fluence.

**Discussion**

An ongoing topic of debate is whether the interspecies excitations involved in OISTR are dipole allowed. We note that selection rules in solids are quite complex, as they are determined by crystal symmetries and often involve states belonging to hybridized bands that exhibit a mixed site- and angular momentum character(52). Furthermore, published calculations of the optical properties of Heuslers do not exclude transitions between bands predominantly associated with differing species(53–55). Moreover, in each of these cases, transitions between hybridized bands(53, 54), or bands from predominantly different species(55), are needed to recreate important features in the characteristic optical response of the material.

Due to limitations on computational power, the TD-DFT framework only simulates one unit cell of Co$_2$MnGa. Therefore, spatial effects such as magnon generation, electron-phonon interactions and superdiffusive spin currents cannot be included. The only demagnetization pathway that can be simulated is individual spin-flips mediated by spin-orbit coupling. For this reason, as discussed above, the theoretical calculations underestimate the amount of demagnetization in the sample, especially on the 100 fs-1 ps timescales where magnon generation becomes prevalent(56–58).

The transient enhancement of the TMOKE signal at the Co peak was quenched at lower fluences than expected. However, at other energies across the Co-edge this was not the case (see Supplementary Materials Fig. S5). We attribute this to strong demagnetization pathways at the Fermi energy, which are not included in the TD-DFT calculations. Further evidence of this is the fact that the Mn demagnetization signal is also underestimated by theory at all probe energies and pump fluences.

The experimental data shown in Fig. 2 show clear and extensive qualitative agreement with the theoretical results that go well beyond what has previously been reported in the literature on this topic. Nevertheless, as noted above, there are a few areas of disagreement. For example, the pre- and post-edge regions of Mn differ in shape. We note that the TD-DFT simulations are, for practical reasons, limited by a few key assumptions. As mentioned above, we only had the computational power to treat one unit cell, and therefore sample imperfections are not considered. The sample is high quality, and grown in the $L2_1$ phase. However, there are many factors that could influence the behavior of a real sample such as interfacial and thin film effects, as well as potential strain from the growth and capping layer. Furthermore, an exact form of the exchange-correlation functional is not known and must be approximated. To account for these approximations in the exchange-correlation functional, the ground state and transient TD-DFT asymmetry spectra are shifted with a rigid blue shift of 0.8 eV and the intensity is scaled with a factor of 1.25 to benchmark the ground state theoretical asymmetry with the experimental measurements.

The time evolution and response functions were calculated using an adiabatic approach. This means that the history and memory dependence of the dynamics were ignored. Therefore, the theoretical dynamics (Fig. 2D) are faster and more intense than the experimental dynamics (Fig. 2B). Depth dependent effects within the sample may also contribute to a slower experimental signal. Furthermore, the experimental pump pulse duration spans from 40 to 55 fs FWHM compared with the theoretically simulated value of 45 fs FWHM. This is because the experimental pulse duration of the laser changes as we tune the central wavelength in order to scan the EUV photon energy across the M-edges.

Most importantly, we have shown here that a simplistic interpretation of the pump-induced changes in the magnetic asymmetry as a change in the magnetic moment is not correct, since same-species optical excitations lead to energy-dependent changes in the asymmetry curve. Fig. 2 shows that the transient enhancement of the TMOKE signal at the Co-edge (both theoretically and experimentally) varies strongly with probing energy. For example, a 5% transient enhancement is measured at 60.4 eV (on resonance) compared with a 14% enhancement measured at 62.5 eV (above edge), as shown in Fig. 2B.

$Co_2MnGa$ has been of particular interest in recent experimental and theoretical studies(*59–64*) due to its topologically non-trivial band structure. Specifically, $Co_2MnGa$ exhibits topological Weyl fermion lines and drumhead surface states(*59*). There has been a recent discussion of the need for ultrafast studies of topological materials(*65*). Measuring transient behaviors could help to characterize and control nodal structures. However, due to the large energy of the exciting laser photons in the experiments presented here (~1.55 eV), we do not have an effective scheme to sensitively measure behavior at Weyl points. In the future, a pump with a lower photon energy could confine more excited electrons within the Weyl points. This could give more insight into the effects of topology and surface states on ultrafast dynamics. Although the topological effects in $Co_2MnGa$ are not addressed in the theoretical analysis in this paper, this work will serve as an important foundation for future ultrafast studies examining the topological nature of this material, and similar systems.

In conclusion, by implementing ultrafast EUV TMOKE at many probing energies across the density of states of Co$_2$MnGa and comparing with TD-DFT, we have established a unified experimental and theoretical framework for understanding complex light-induced spin dynamics on very short timescales (<100 fs). A prominent finding of this investigation is that the pump-induced change of the asymmetry of the EUV TMOKE signal can vary substantially from the changes of the magnetic moment. In addition, we have distinguished intrasite, intersite and spin-flip contributions to the transient TMOKE signal and their implications on the interpretation of the transient signal. We have also made the first fluence-dependent measurements of spin transfer effects and observe that the resonant enhancements of Co are maximized at surprisingly low fluences (1.0 mJ/cm$^2$). The differing fluence and energy dependent behaviors across the Co-edge demonstrate that experiments claiming to observe OISTR need to probe across the entire edge to disentangle the contribution of different microscopic processes to the magnetic asymmetry.

Our extensive experiment and theory datasets have allowed us to identify key regions of theoretical and experimental agreement along with areas for future improvement. The theoretical simulations and experimental measurements jointly demonstrate large energy dependent spin transfer signals. Thus, this work sets a high standard for theoretical and experimental agreements for ultrafast spin dynamics in alloys and provides insight into competing microscopic mechanisms.

**Experimental Design**

Ultrafast pulses from a regenerative Ti:Sapphire amplifier are split and simultaneously used as a ~800 nm pump and EUV probe produced using HHG. The energy of the probe was tuned using a combination of three different methods: firstly, by tuning the central wavelength and bandwidth of the seed pulse into the amplifier; secondly, by applying gain flattening filters to the seed pulse to redshift the resultant amplified pulse; and thirdly, by applying chirp (second order dispersion) to the pulse to preferentially involve certain wavelengths in the high harmonic process. We note that small energy changes in the driving laser result in relatively much larger changes to the probe energy due to the additive nature of HHG. Since we do not have independent control over the pump pulse we must consider how this engineering of the driving laser impacts the pump parameters. The pump pulse compressibility is impacted when we detune from the amplifier's central energy using the three techniques described above. This results in a range of pump pulse durations from 40 fs to 55 fs depending on the required probe energy. Furthermore, the brightest wavelength of the pump exists in a range from 775 nm-805 nm depending on the required probe energy. Since these changes to the pump pulse are minimal, for the purpose of theory and interpretation, we use an 800 nm (1.55 eV) pump with a 45 fs duration for all calculations. However, we note that the experimental range of pump energies from 1.54 eV to 1.60 eV will have some impact on the allowable pump transitions, especially across the bandgap in the minority channel. The experimental time zero was determined by detecting any change in the magneto-optical signal that indicates the onset of the pump pulse. The entire harmonic comb is measured simultaneously by the CCD chip. This means that time-dependent data at both the Mn-edge, Co-edge, as well as all the energies in between, can be considered when determining time zero. A change in the magneto-optical signal (i.e. time zero) is defined as an increase or decrease (relative to the ground state) which falls outside the ±1 standard deviation error bars for one or more of the harmonic orders. The theoretical time zero was chosen to maximize correspondence with the experimental data. The definition of the theoretical time zero relative to the pump pulse is shown in the Supplementary Materials (Fig. S7). More specific details of the X-MATTER beamline are described in our recent design paper(*48*), and a brief schematic appears in the Supplementary Materials (Fig. S1).

**Statistical Analysis**

Time points in the pump-probe experiments were measured in a random order to prevent systematic errors. The randomized set of time points were measured repeatedly until a sufficient signal-to-noise ratio was obtained. Source noise from the high harmonic generation was monitored and digitally cancelled using the methods described in Johnsen *et al.*(*48*). Each experimental data point represents the arithmetic mean of repeated measurements. The error bars represent the standard deviations of each measurement.

**Theoretical Framework**

TD-DFT maps the time-dependent many-body interacting problem to an equivalent non-interacting Kohn Sham system in an effective potential that reproduces exactly the same density as the interacting one. The time-dependent Kohn-Sham (TDKS) Hamiltonian can be written as: observed.

$$\left[ \frac{1}{2}\left(-i\nabla + \frac{1}{c}\mathbf{A}_{ext}(t)\right)^2 + v_s(\mathbf{r},t) + \frac{1}{2c}\sigma\cdot\mathbf{B}_s(\mathbf{r},t) + \frac{1}{4c^2}\sigma\cdot\left(\nabla v_s(\mathbf{r},t)\times -i\nabla\right)\right]\psi_i(\mathbf{r},t) = \frac{\partial \psi_i(\mathbf{r},t)}{\partial t}$$

(1)

where $c$ is the speed of light, $\sigma$ is the Pauli matrix, and $\mathbf{B}_s(\mathbf{r},t)$ is the effective Kohn-Sham (KS) magnetic field. $\mathbf{B}_s(\mathbf{r},t)$ is the sum of two terms: $\mathbf{B}_s(\mathbf{r},t) = \mathbf{B}_{ext}(t) + \mathbf{B}_{XC}(\mathbf{r},t)$, where $\mathbf{B}_{ext}(t)$ is the magnetic field of the external laser pulse and $\mathbf{B}_{XC}(\mathbf{r},t)$ is the exchange-correlation (XC) induced exchange splitting. The KS effective potential, $v_s(\mathbf{r},t)$, is a sum of three terms $v_s(\mathbf{r},t) = v_{ext}(\mathbf{r},t) + v_H(\mathbf{r},t) + v_{xc}(\mathbf{r},t)$ where $v_{ext}(\mathbf{r},t)$ is the external potential, $v_H(\mathbf{r},t)$ is the Hartree potential, and $v_{xc}(\mathbf{r},t)$ is the exchange-correlation potential. The last term of Eq. (1) is the spin-orbit coupling (SOC) term in its generic form, and $\psi_i(\mathbf{r},t)$ is the two-component Pauli spinor. The external laser pulse is treated in the dipole approximation with a vector potential $\mathbf{A}_{ext}(t)$. The atomic units (with $\hbar = e = m = 1$) are adopted in Eq. (1). observed.

Eq. (1) is propagated in time under the influence of the pump laser using the implementation of the ELK code(*66*).

**Transient Response Function**

A mixed scheme between the time evolution (TE) of Eq. 1 and the linear response is employed to calculate the response function at a given time, as introduced by Dewhurst *et al.*(*67*). The non-interacting (KS) response function is calculated as(*68*):

$$\chi_{KS}(\mathbf{r},\mathbf{r}',\omega) = \lim_{\eta \to 0} \sum_{i=1}^{\infty} \sum_{j=1}^{\infty} (n_i - n_j) \frac{\psi_i^*(\mathbf{r})\psi_j(\mathbf{r})\psi_i(\mathbf{r}')\psi_j^*(\mathbf{r}')}{\omega - \epsilon_j + \epsilon_i + i\eta} \quad (2)$$

where $\psi_i$ are the KS orbitals, $\epsilon_i$ are the eigenvalues, and $n_i$ are the occupation numbers. The interacting response function, $\chi$, is related to the non-interacting one, $\chi_{KS}$, through a Dyson equation of the form:

$$\chi(\omega) = \chi_{KS}(\omega) + \chi_{KS}(\omega)(v + f_{xc}(\omega))\chi(\omega) \quad (3)$$

where $v$ is the bare Coulomb interaction and $f_{XC}$ is the exchange-correlation kernel(*69*). Finally, the dielectric tensor is calculated from the response function as (*69*):

$$\epsilon_{ij}^{-1}(\omega) = \delta_{ij} + v\chi_{ij}(\omega) \quad (4)$$

where $\epsilon_{ij}(\omega)$ is the dielectric tensor.

We emphasize that our approach for calculating the dielectric tensor of the transient state is substantially different from the state-blocking method in Ref. (*70*) and also different from Ref. (*67*) and their subsequent work in Ref. (*10*, *13*), and (*42*) where only the transient occupations from the time evolution of Eq. 1 are used to evaluate Eq. 2 but the excitation energies and the KS states appearing in Eq. 2 are taken from the ground state (GS). We note that fixing the excitation energies to the GS ones implies a rigid band structure and therefore any dynamical shifts of the transient spectra as reported by L-edge measurements in Ref. (*44*) cannot be captured. Furthermore, using the GS KS orbitals implies that the valence orbital shape and its symmetry remain the same even upon the application of the pump laser. However, the application of the pump laser reduces the symmetry of the system and hence this changes the transition probabilities from the core (3p) to the valence (3d) states. Therefore, we find using full transient quantities resulting from the time evolution of Eq. 1 to evaluate the KS response function is more plausible, leading to a very good agreement with the experimental observables in the TMOKE experiment.

The energy of the static and transient TMOKE spectra are shifted with a rigid blue shift of 0.8 eV and the intensity is scaled with a scaling factor of 1.25 to match the experiment. These discrepancies can be attributed primarily to many-body effects of the core hole correlations which are only roughly described in Kohn-Sham DFT due to the approximation of the exchange-correlation functional with the local spin density approximation (LSDA)(*71*).

We accounted for the experimental geometry by solving the Fresnel equations numerically using the calculated dielectric tensor as detailed in Ref. (*71–73*). The non-magnetic contribution from the capping layer is treated as described in Ref. (*72*).

**Projected Density of States**

The occupation of the KS transient state, $\eta(\epsilon,t)$, is calculated by first projecting the TDKS state, $\psi(r,t)$, on the time independent, ground state, $\phi(r)$, via:

$$P_{ij}^{k}(t) = \int d^{3}r \phi_{ik}^{*}(r)\psi_{jk}(r,t) \tag{5}$$

We obtain the time-dependent occupation projected on the ground state by summing the square of the projection $P_{ij}^{k}(t)$ over all TDKS states, weighted by the occupation number, $n_{jk}$, as given by:

$$w_{ik}(t) = \sum_{j} n_{jk} \left| P_{ij}^{k}(t) \right|^{2} \tag{6}$$

Finally, the time-dependent projected DOS, $\eta(\epsilon,t)$, is evaluated according to:

$$\eta(\epsilon,t) = \sum_{i}^{\infty} \int_{BZ} \delta(\epsilon - \epsilon_{ik}) w_{ik}(t) \tag{7}$$

where $\epsilon_{ik}$ is the $i^{th}$ Kohn-Sham energy eigenvalue. The changes in $\eta(\epsilon,t)$ at 70 fs with respect to the ground state ($t$ = 0) are presented in Figs. 3A and 3B.

**Computational parameters**

The calculations are performed for a unit cell of Co$_2$MnGa in the L2$_1$ phase. The calculations of this system were performed in a fully *ab initio* and non-collinear fashion with 10 × 10 × 10 k-points in the Brillouin zone. The GS evolved in time using Eq. (1) with a time step of 2.4 attoseconds(*66*). We used the local spin density approximation (LSDA) and adiabatic LSDA as an exchange-correlation functional for the ground state and time evolution calculations respectively. A smearing width of 0.027 eV was used for the ground state and for time propagation, but for the response function calculations, a smearing width of 0.27 eV was used. A laser pulse with a wavelength of 800 nm, fluence of 2.2 mJ/cm², and full width at half maximum (FWHM) of 45 fs was allowed to interact with the electronic subsystem and the response to this external field was followed for over 100 fs.

**Sample Growth**

20-nm-thick (001)-oriented Co$_2$MnGa film was grown on (001)-oriented MgO single-crystal substrate using a BESTEC ultra-high vacuum magnetron sputtering system. Before deposition, the chamber was evacuated to a base pressure of less than 6 × 10⁻⁹ mbar, whereas the process gas (Ar 5 N) pressure was 3 × 10⁻³ mbar. The target-to-substrate distance was 20 cm, and the substrate was rotated at 20 rpm to ensure homogeneous growth. We used Co (5.08 cm), Mn (5.08 cm) and $Mn_{50}Ga_{50}$ Si (5.08 cm) sources in confocal geometry by applying 34 W, 6W and 20 W DC power, respectively. The films were grown at 640 °C followed by an additional 30 min in-situ post annealing at the same temperature and capped with an amorphous 3 nm-thick Si film deposited at room temperature to prevent the oxidation of the film. More information on the sample growth and properties are available in this study(*73*).

**Funding Acknowledgements**

The JILA team gratefully acknowledges support for this research from the Department of Energy Office of Basic Energy Sciences X-Ray Scattering Program under Award No. DE-SC0002002. P.C.J. and A.G. were supported by the NSF GRFP.

O.G. gratefully acknowledges financial support from the Strategic Research Council (SSF), grant ICA16-0037 and the Swedish Research Council (VR), grant number 2019-03901.

A.D. gratefully acknowledges financial support from the Swedish Research Council (VR), grant numbers 2016-05980 and 2019-05304; and the Knut and Alice Wallenberg foundation (KAW), grant numbers 2018.0060, 2021.0246, and 2022.0108.

O.E. gratefully acknowledges financial support from the European Research Council via Synergy Grant 854843 – FASTCORR; eSSENCE; STandUPP; the Swedish Research Council (VR); and the Knut and Alice Wallenberg foundation (KAW), grant numbers 2018.0060, 2021.0246, and 2022.0108. GRFP.

A.D. and O.E. gratefully acknowledge the Wallenberg Initiative Materials Science for Sustainability (WISE), funded by the Knut and Alice Wallenberg Foundation.

M.E., A.D., O.G, and O.E. gratefully acknowledge computational resources provided by the National Academic Infrastructure for Supercomputing in Sweden (NAISS) and the Swedish National Infrastructure for Computing (SNIC) at NSC and PDC, partially funded by the Swedish Research Council through grant agreements no. 2022-06725 and no. 2018-05973.

**Author Contributions**

S.A.R., P.C.J., A.G., N.L., H.C.K. and M.M.M. conducted the experiments and analyzed the experimental data. M.F.E. performed the theoretical calculations under the supervision of O.G., O.E., and A.D. The sample was prepared and characterized by A.M., E.L. and C.F. All authors contributed to writing the manuscript.

**Competing Interests**

H.C.K. and M.M.M. have a financial interest in a laser company, KMLabs, that produces the lasers and HHG sources used in this work. H.C.K. is partially employed by KMLabs. All other authors declare they have no competing interests.

# Supplementary Materials

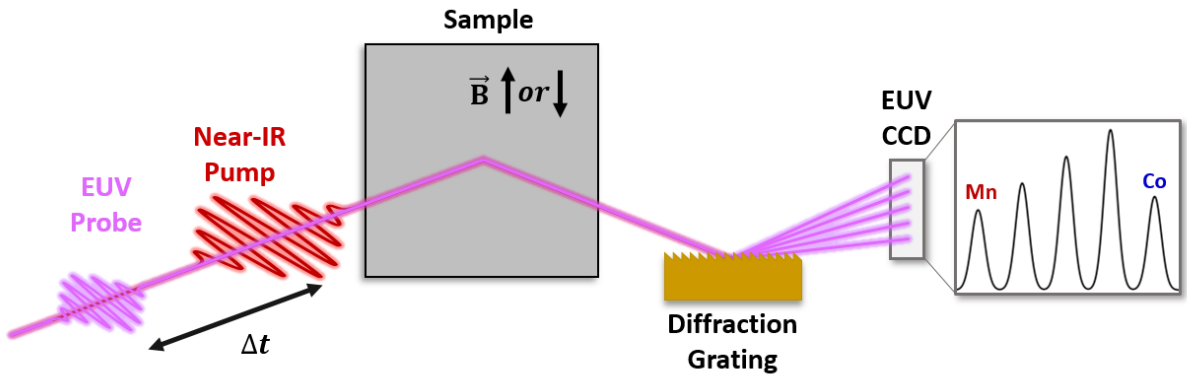

**Fig. S1. Experimental design.** The magnetized sample is excited with a near-IR pump pulse followed by an EUV probe pulse after a time delay $\Delta t$. The EUV probe contains a comb of energies produced by high harmonic generation. The harmonic comb is spectrally dispersed using a diffraction grating then detected with a CCD camera. Different harmonic energies are resonant with the M-edges of Mn and Co. TMOKE measurements are made by comparing the intensity of light reflected from the sample with two different sample magnetization directions (up and down) as shown in the figure.

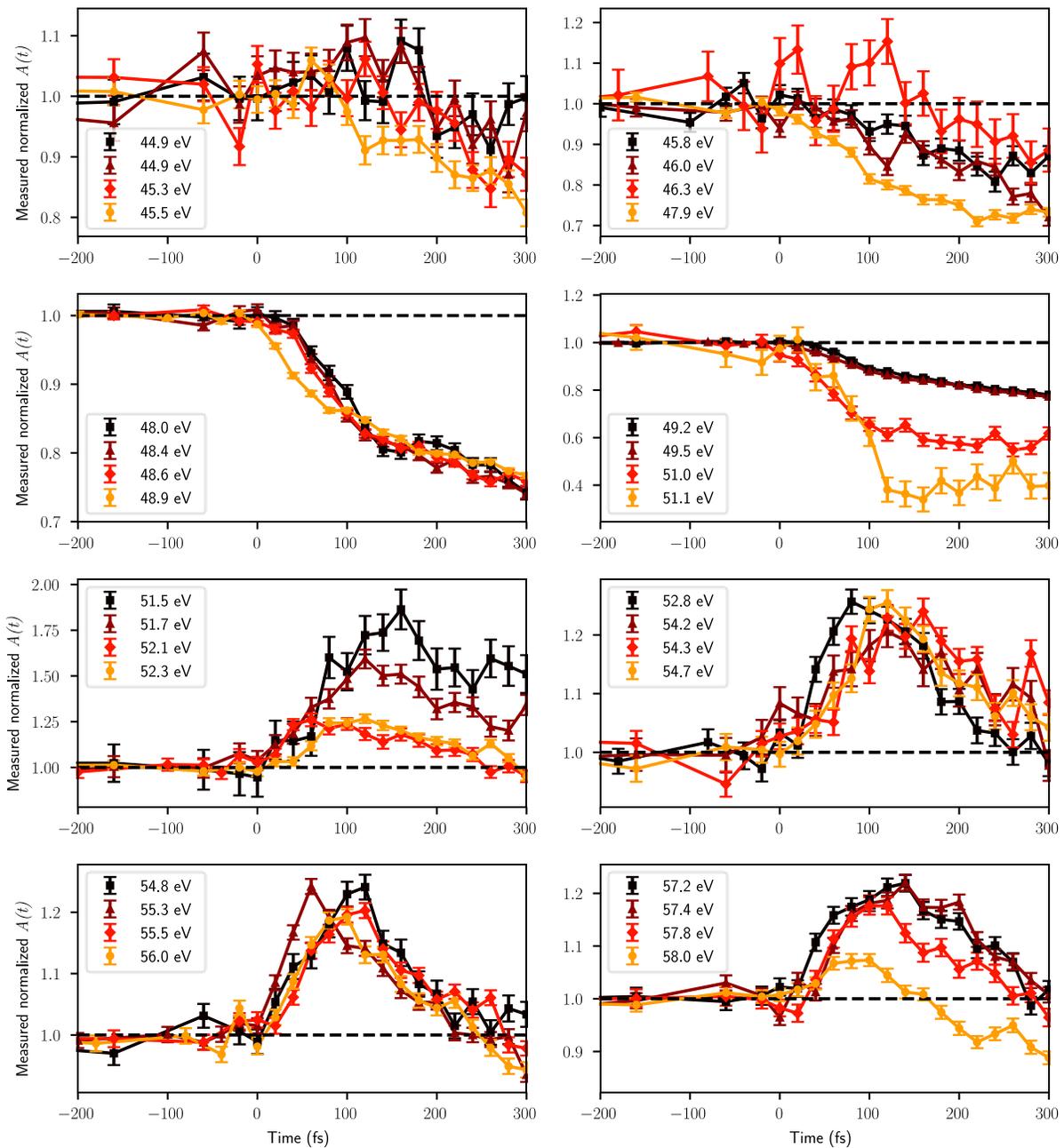

**Fig. S2. Energy dependent transient magnetic asymmetry with probing energies from 44.9 eV to 58.0 eV.** The peak of the Mn asymmetry occurs at 49.2 eV. The probing energy region from 51.5 eV to 55.0 eV lies between the Co and Mn edges. Signals at these probing energies see a combination of high energy Mn states as well as low energy Co states. We do not draw any conclusions from the enhancements seen in this region due to their strongly mixed nature.

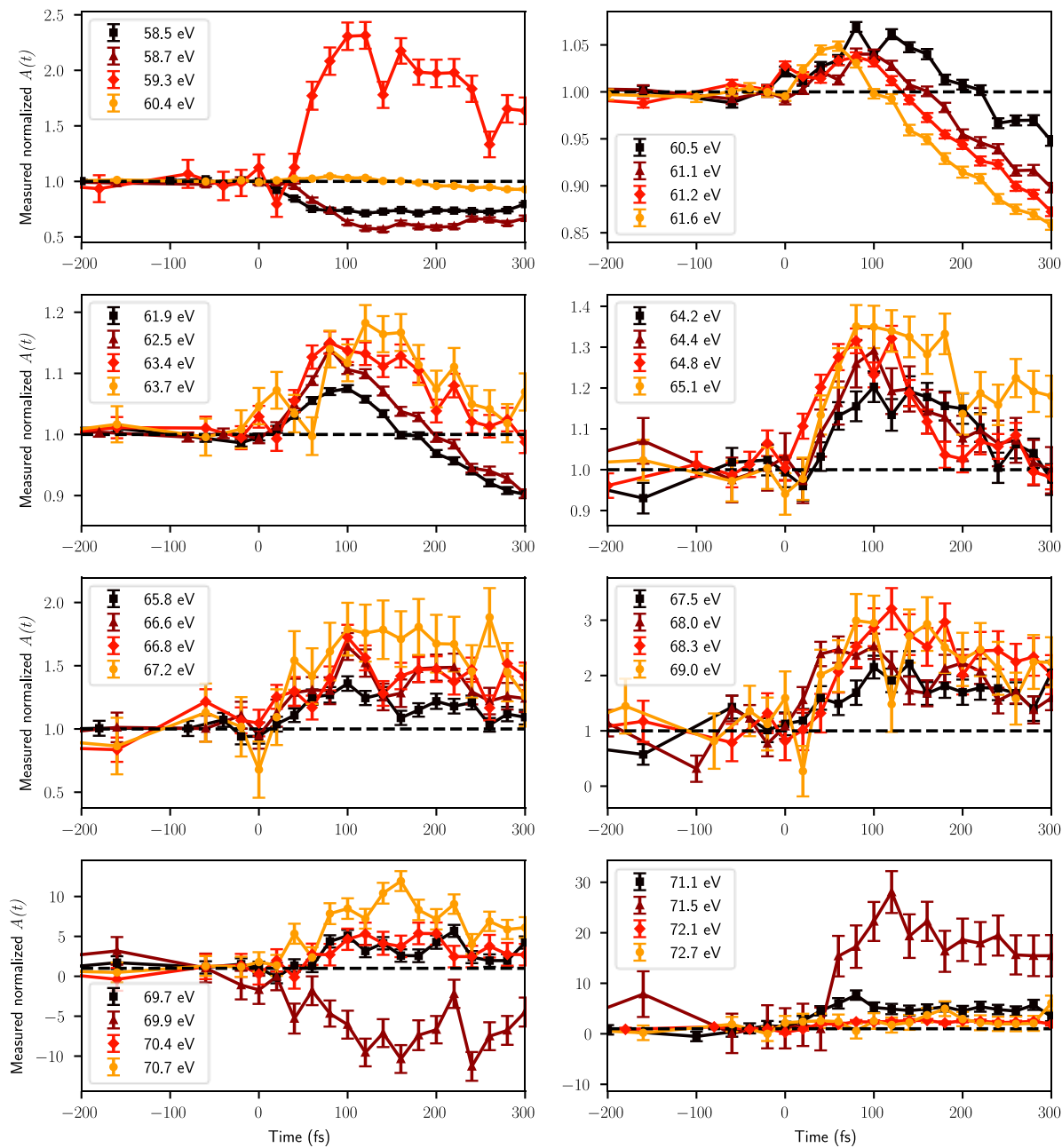

**Fig. S3. Energy dependent transient magnetic asymmetry with probing energies from 58.5 eV to 72.7 eV.** The peak of the Co asymmetry occurs at 60.4 eV.

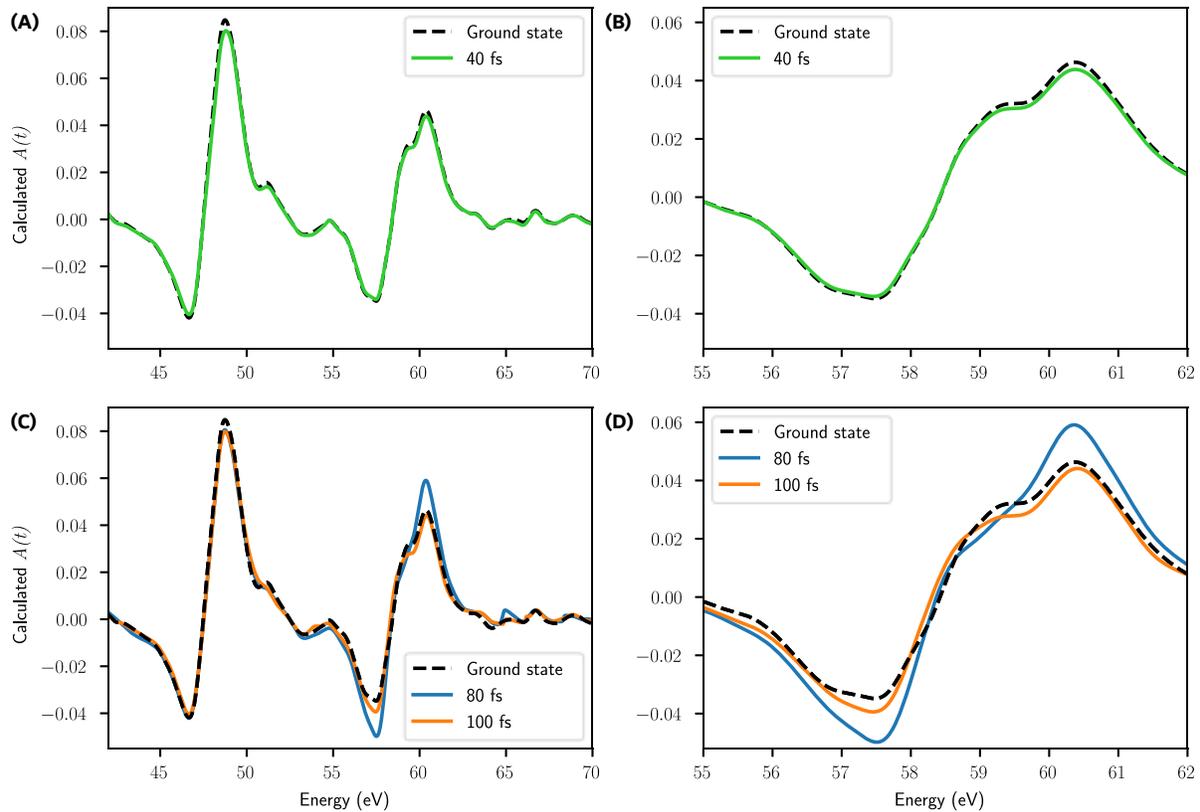

**Fig. S4. Theoretical asymmetry curves at 0 fs, 40 fs, 80 fs and 100 fs following laser excitation from the pump pulse.** (A) The asymmetry at 40 fs plotted across both the Co and Mn edges and (B) zoomed in on the Co-edge. At 40 fs, we see a reduction in asymmetry at the Mn and Co resonant peaks due to spin-flips. (C) The asymmetry at 80 fs and 100 fs and (B) zoomed in at the Co-edge. At 80 fs, we see enhancements across the Co-edge due to spin transfers. At 100 fs, the spin transfer excitations have mostly decayed and spin-flips begin to dominate once again.

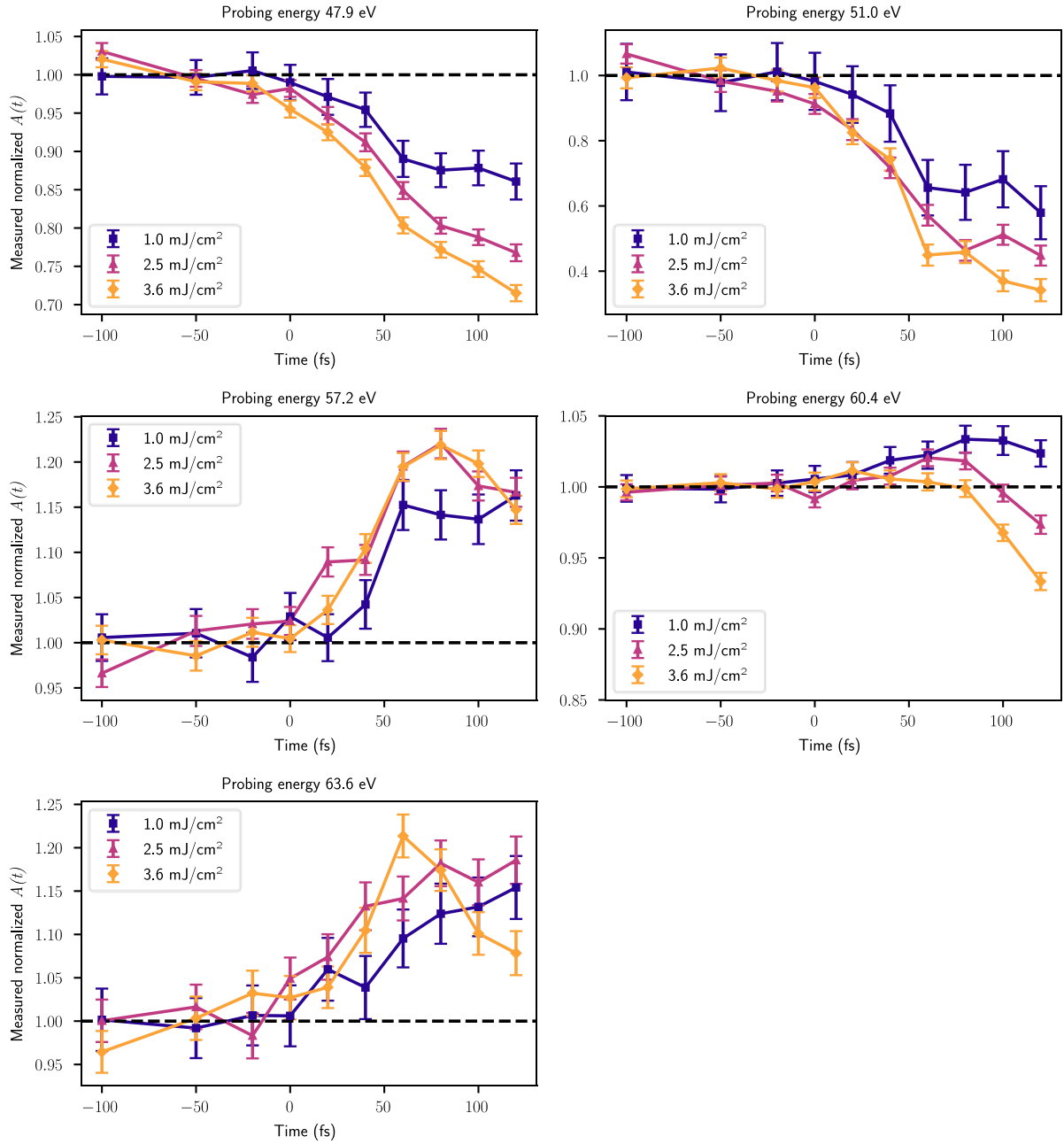

**Fig. S5. Fluence dependent transient magnetic asymmetry measurements with five different probing energies from 47.9 eV to 63.6 eV**. Around the Mn-edge, (i.e. 47.9 eV and 51.0 eV), demagnetization dominates at all fluences. Above and below the Co-edge at 63.6 eV and 57.2 eV, spin transfer dominates and the signal in enhanced. At the Co-edge, 60.4 eV, there is fluence dependent competition between spin transfer and ultrafast demagnetization.

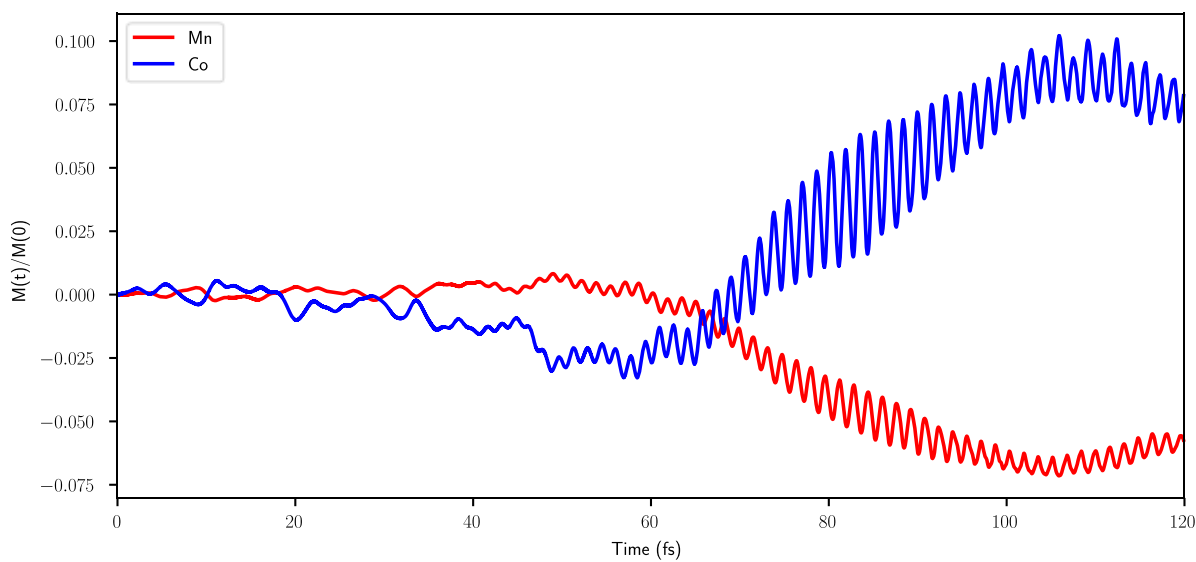

**Fig. S6. The simulated transient changes in magnetic moment of Co and Mn with a pump fluence of 8.4mJ/cm$^2$.** Here, a large calculated OISTR effect of 10% for Co is depicted for an 8.4 mJ/cm$^2$ pump fluence. In comparison, the Co moment only increases by ~2.5% for 2.2 mJ/cm$^2$ pumping. However, we note that the simulation does not include all effects that would contribute to demagnetization of Co such as: magnon generation, electron-phonon interactions, and superdiffusive spin currents. For this reason, we expect that this is an overestimation of the strength of the Co moment increase.

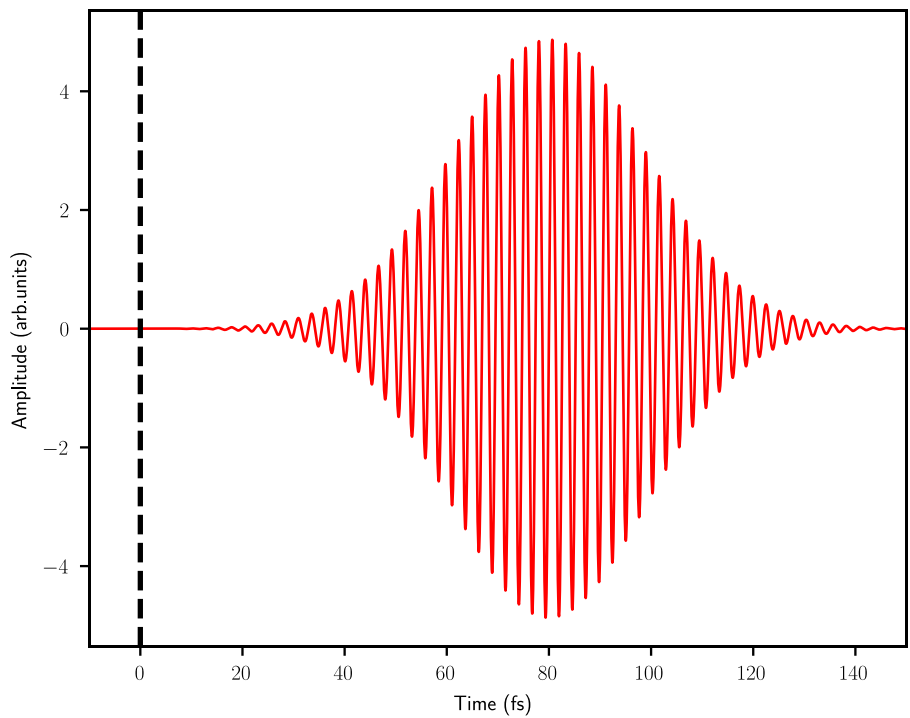

**Fig. S7. The theoretical definition of *t*=0 relative to the time dependent amplitude of the simulated incident pump pulse.** The theoretical time zero was chosen to maximize agreement with the experimentally determined time zero.

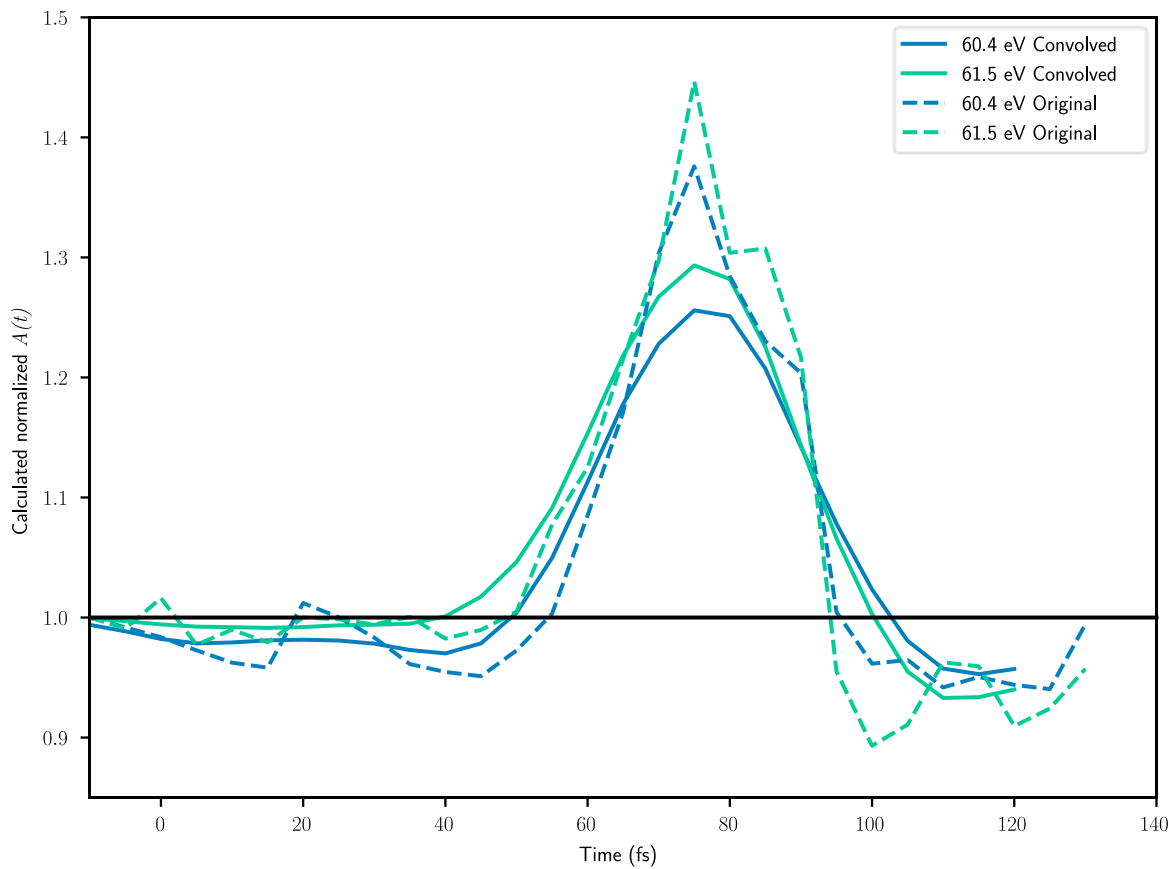

**Fig. S8. The calculated theoretical magnetic asymmetry dynamics with and without convolving with a measurement probe.** Dashed lines: the instantaneous magnetic asymmetry calculated for two different probing energies: 60.4 eV and 61.5 eV in steps of 5 fs. Solid lines: the same theoretical data convolved with a 25 fs FWHM probe pulse as described in the main text Fig. 2(D).